\documentclass[prb,twocolumn,superscriptaddress,floatfix]{revtex4}
\usepackage{graphicx}
\usepackage{booktabs}

\begin{document}

\title{Raman Spectroscopic Features of the Neutral Vacancy in Diamond from \textit{Ab Initio} Quantum-mechanical Calculations}

\author{J. Baima}
\affiliation{Dipartimento di Chimica, Universit\`{a} di Torino, Via Pietro Giuria 5, 10125 Torino, Italy}
\affiliation{Centro Inter-dipartimentale \textquotedblleft Nanostructured Interfaces and Surfaces\textquotedblright, Universit\`{a} di Torino, Via Gioacchino Quarello 15/A, 10135 Torino, Italy}
\author{A. Zelferino}
\affiliation{Dipartimento di Fisica, Universit\`{a} di Torino, Via Pietro Giuria 1, 10125 Torino, Italy}
\author{P. Olivero}
\affiliation{Dipartimento di Fisica, Universit\`{a} di Torino, Via Pietro Giuria 1, 10125 Torino, Italy}
\affiliation{Centro Inter-dipartimentale \textquotedblleft Nanostructured Interfaces and Surfaces\textquotedblright, Universit\`{a} di Torino, Via Gioacchino Quarello 15/A, 10135 Torino, Italy}
\author{A. Erba}
\author{R. Dovesi }
\affiliation{Dipartimento di Chimica, Universit\`{a} di Torino, Via Pietro Giuria 5, 10125 Torino, Italy}
\affiliation{Centro Inter-dipartimentale \textquotedblleft Nanostructured Interfaces and Surfaces\textquotedblright, Universit\`{a} di Torino, Via Gioacchino Quarello 15/A, 10135 Torino, Italy}

\begin{abstract}

Quantum-mechanical {\it ab initio} calculations are performed to elucidate the vibrational spectroscopic features of a common irradiation-induced defect in diamond, \textit{i.e.} the neutral vacancy. Raman spectra are computed analytically through a Coupled-Perturbed-Hartree-Fock/Kohn-Sham approach as a function of both different defect spin states and defect concentration. The experimental Raman features of defective diamond located in the $400 - 1300$ cm$^{-1}$ spectral range, \textit{i.e.} below the first-order line of pristine diamond at 1332 cm$^{-1}$, are well reproduced, thus corroborating the picture according to which, at low damage densities, this spectral region is mostly affected by non-graphitic sp$^3$ defects. No peaks above 1332 cm$^{-1}$ are found, thus ruling out previous tentative assignments of different spectral features (at 1450 and 1490 cm$^{-1}$) to the neutral vacancy.
The perturbation introduced by the vacancy to the thermal nuclear motion of carbon atoms in the defective lattice is discussed in terms of atomic anisotropic displacement parameters (ADPs), computed from converged lattice dynamics calculations.
  
\end{abstract}

\maketitle

\section{Introduction}
\label{intro}

Since many years, the investigation of native and radiation-induced point-defects in semiconductors has attracted an ever-increasing interest in both theoretical and experimental studies. This is particularly true in the case of diamond, a wide-bandgap material characterized by well-known extreme physical properties (high Young's modulus and thermal conductivity, broad transparency range, high carriers mobility, etc.) with attractive applications in different fields, ranging from microelectromechanical systems to heatsinks, laser windows, particle detectors, etc.\cite{1:jackman2003,2:nemanich2014} Indeed, the presence of defects in the crystal structure of diamond has a dramatic effect on its physical properties, from a structural,\cite{3:hoffman1992} optical, \cite{4:lagomarsino2012} and electronic\cite{5:prins1985} point of view. A rigorous understanding of the physical effects of different typologies of defects is therefore of paramount importance in diamond science and technology.

Despite the large number of studies on the subject, \cite{6:kalish1999-1,7:kalish1999-2,8:twitchen1999,9:lai2002,10:morono2007,11:amekura2008} the defect formation mechanisms in diamond are still far from having been exhaustively explored and understood. This is especially due to the peculiar meta-stability of its crystalline structure, in which both sp$^{3}$ and graphitic-like sp$^{2}$ chemical bonds give rise to an unusual variety of different point-defects and related complexes. In this respect, Raman spectroscopy is the ideal experimental technique to study different carbon allotropes due to their characteristic vibrational features,\cite{12:ferrari2000,13:ferrari2001} and it has thus emerged as a prominent technique to investigate defect formation in irradiated diamond. \cite{14:jamieson1995,15:hunn1995,16:prawer1998,6:kalish1999-1,7:kalish1999-2,17:orwa2000,11:amekura2008,18:brunetto2004,19:olivero2006,20:prawer2008,21:bergman2009} 

The Raman spectrum of pristine diamond consists of a single, sharp Raman peak at 1332 cm$^{-1}$, corresponding to the first-order scattering with triply-degenerated TO(X) phonons of F$_{2g}$ symmetry.  The damaged crystal, on the contrary, is characterized by several additional features, which have been attributed to different types of sp$^{2}$ and sp$^{3}$ defects. The most prominent features observed at higher Raman frequencies with respect to the first-order line are located at about $1450$, $1490$, $1630$ and $1680$ cm$^{-1}$. The $1630$ and $1680$ cm$^{-1}$ features are commonly attributed to sp$^{2}$ defects such as the ``dumb-bell'' split-interstitial defect.\cite{16:prawer1998,6:kalish1999-1}
On the other hand, different tentative attributions have been formulated for the $1450$ and $1490$ cm$^{-1}$ peaks, involving both vacancy\cite{16:prawer1998,20:prawer2008} and intrinsic/nitrogen interstitial defects.\cite{24:woods1984, 25:linchung1994} At lower Raman frequencies, two broad bands are measured in defective diamond, in the $400-1000$ cm$^{-1}$ and $1000-1300$ cm$^{-1}$ ranges, respectively.\cite{14:jamieson1995, 15:hunn1995, 17:orwa2000} These vibrational bands are rather articulated and their attribution is unclear. \cite{22:zaitsev2001} As these broad bands qualitatively resemble the vibrational density-of-states of pristine diamond,\cite{23:wang1993} they have been tentatively attributed to sp$^{3}$ defects. \cite{16:prawer1998} Moreover, the first-order Raman peak itself is subjected to both broadening and red-shifting when damage is introduced in the crystal,\cite{14:jamieson1995, 17:orwa2000} an effect which can be interpreted by considering that the peak frequency in the first-order Raman spectrum primarily depends on the bond length and strength: as the inter-atomic distance increases in defective diamond, the Raman shift frequency lowers.

Due to recent advances in the theoretical development and computational implementation of quantum-mechanical fully-analytical methods for the {\it ab initio} evaluation of Raman spectra of solids,\cite{maschio_2013_1,maschio_2013_2,CRYSTAL14PAP} computational Raman spectroscopy has become an effective complementary tool in the interpretation of experimental spectra. As a first step towards the characterization of the Raman spectral features of defective diamond, in this study Raman spectra are computed as a function of concentration and different spin states of neutral vacancies, and compared to previous experimental results. The thermal nuclear motion of defective diamond is also characterized by means of converged lattice dynamical calculations and anisotropic displacement parameters (ADPs). 

Within density-functional-theory (DFT) the global hybrid functional B3LYP~\cite{Becke,LYP,B3LYP} is adopted, which includes 20\% of non-local exact Hartree-Fock exchange and is known to outperform other classes of functionals in the description of vibration properties of most solids~\cite{ugl:1,cat:1,bar:1} and in the accurate description of spin states.\cite{moreira2004spin,munoz2004spin,wojdel2008spin} 

%A global hybrid functional as B3LYP~\cite{Becke,LYP,B3LYP} is adopted within density-functional-theory (DFT), which includes 20\% of non-local exact Hartree-Fock exchange and is known to outperform other classes of functionals in the description of vibration properties of most solids,\cite{dovesi2015vibration,ugl:1,cat:1,bar:1} as well as 

\section{Computational Methods and Details}
\label{sec:meth}

\subsection{Spectroscopic Features}

Harmonic phonon frequencies, $\omega_p$, at the $\Gamma$ point ({\it i.e.} at the center of the first Brillouin zone, FBZ) are obtained from the diagonalization of the mass-weighted Hessian matrix of the second energy derivatives with respect to atomic displacements $u$:~\cite{319,freq2,Aragonite}
\begin{equation}
\label{eq:hess1}
W^{\Gamma}_{a i,b j} = \frac{H^{\bf 0}_{a i,b j}}{\sqrt{M_a M_b}} \quad \textup{with} \quad H^{\bf 0}_{a i,b j} = \left( \frac{\partial^2 E}{\partial u_{a i}^{\bf 0} \partial u_{b j}^{\bf 0}} \right)\; ,
\end{equation}
where atoms $a$ and $b$ (with atomic masses $M_a$ and $M_b$) in the reference cell, {\bf 0}, are displaced along the $i$-th and $j$-th Cartesian directions, respectively. First order derivatives are computed analytically, whereas second order derivatives are obtained numerically, using a two-point formula.

The Raman intensity of the Stokes line of a phonon mode $Q_{p}$, 
active due to the $\alpha_{ii^\prime}$ component of the polarizability tensor $\alpha$, can be expressed as:
\begin{equation}
\label{eq:eqint}
I^{p}_{ii^\prime}\propto \left(\frac{\partial\alpha_{ii^\prime}}{\partial Q_{p}}\right)^{2}\; .
\end{equation}
The relative Raman intensities of the peaks are computed analytically from the solution of second-order Coupled-Perturbed-Hartree-Fock/Kohn-Sham (CPHF/KS) equations,~\cite{mauro1,mauro2} exploiting a scheme recently implemented in the \textsc{Crystal14} program.~\cite{maschio_2013_1,maschio_2013_2} The Raman spectrum is then computed by considering the transverse optical (TO) modes and by adopting a pseudo-Voigt functional form: a linear combination of a Loren\-tzi\-an and a Gaussian curve with FWHM of 8 cm$^{-1}$. Raman intensities are normalized so that the largest value is conventionally set to 1000 a.u. 

%Integrated intensities for infrared (IR) absorption are computed for each mode by means of the mass-weighted effective mode Born charge vector, evaluated through an analytical approach.~\cite{maschio2,irletter} Both schemes (for Raman and IR spectra) are based on the solutions of first- and second-order Coupled-Perturbed-Hartree-Fock/Kohn-Sham (CPHF/KS) equations.~\cite{mauro1,mauro2}  

\subsection{Phonon Dispersion}
\label{subsec:phon}

The calculation of other vibrational properties of solids (such as thermodynamic quantities or thermal nuclear motion indices) is a more demanding task as it implies
knowledge of the phonon dispersion inside the full FBZ.~\cite{PyrGro} Beside
$\mathbf{W}^{\Gamma}$, in this case a set of dynamical matrices,
$\mathbf{W^k}$, needs to be formed for a set of wavevectors
$\mathbf{k} =\sum_{i} \frac{\kappa_i}{L_i}\, \mathbf{b}_i$ expressed as
linear combinations of reciprocal lattice basis vectors $\mathbf{b}_i$
with fractional coefficients. Such coefficients are determined by the shrinking factors $L_i$, as well as by integers $\kappa_i$
ranging from 0 to $L_i -1$, thus including $\Gamma$ and
points within the FBZ. Phonons at $\mathbf{k}$ points other than $\Gamma$ can be obtained by the
\textit{direct method},~\cite{parlinski97,phonopy,NucECD} which requires the
construction of supercells (SC) of the original unit cell:
\begin{equation} 
\label{eq:dynmat}
W^{\mathbf{k}}_{a i,b j} = \sum_{\mathbf{g}\in SC}\, 
\frac{H^{\mathbf{g}}_{a i,b j}}{\sqrt{M_a M_b}} 
\, e^{\imath \mathbf{k}\cdot \mathbf{g}} \; .
\end{equation} 
Indeed, equation (\ref{eq:dynmat}) shows that each dynamical matrix in the FBZ is
obtained by Fourier-transforming the Hessian matrices, $\mathbf{H^g}$, for an
adequate set of real-space lattice vectors $\mathbf{g}$. Such lattice vectors
$\mathbf{g} \, = \, \sum_{i} l^{g}_{i}\, \mathbf{a}_i$ can be expressed in terms
of the real lattice basis vectors $\{ \mathbf{a}_i \}$ through the integer
coefficients $l^{g}_{i}$; they are all contained in the real-space SC, whose
size and shape are determined by parameters $L_i$. At variance with equation
(\ref{eq:hess1}), each matrix element $H^{\mathbf{g}}_{a i,b j} =
{\partial^2 E}/({\partial u^{\mathbf{0}}_{a i} \partial
u^{\mathbf{g}}_{b j}})$ refers to a displacement of atom $b$ in cell
$\mathbf{g}$ inside the SC along the Cartesian direction $j$, along with all
its images throughout the superlattice generated by the SC. $L_i$ parameters are the same
both in the real and the reciprocal space, so as to maintain a one-to-one
matching between $\mathbf{g}$ vectors in the SC and sampled $\mathbf{k}$ points. From the diagonalization of the dynamical matrices, the normal modes and corresponding vibration frequencies are
sampled over the entire FBZ:
\begin{equation}
\label{eq:diag}
({\bf U}^{\bf k})^{\dagger}\,{\bf W}^{\bf k}\,{\bf U}^{\bf k}\;=\;
{\bf \Lambda}^{\bf k} \quad \textup{with} \quad ({\bf U}^{\bf k})^{\dagger}{\bf U}^{\bf k} = {\bf I}\; .
\end{equation}
The elements of the diagonal ${\bf \Lambda}^{\bf k}$ matrix provide the {\it vibrational frequencies}, $\nu_{{\bf k}p}=\sqrt{\lambda_{{\bf k}p}}$ (atomic units are adopted), while the columns of the ${\bf U}^{\bf k}$ matrix contain the corresponding {\it normal coordinates} $\bf{u}^{\bf k}_{p}$ (whose elements are ${u}^{\bf k}_{ia,p}$, being $ia$ a combined index running from 1 to 3$M$, where $M$ is the number of atoms per cell). To each ${\bf k}$-point in the first Brillouin zone, 3$M$ harmonic oscillators (\textit{i.e.} phonons) are associated, which are labeled by a phonon band index $p$ ($p= 1, \dots , $3$M$).

\subsection{Atomic Thermal Motion}
\label{subsec:adp}

Atomic anisotropic displacement parameters (ADPs) are commonly adopted to analyze the mean square atomic displacements due to thermal nuclear motion, particularly in the field of X-ray diffraction. Due to zero-point and thermal motion, each atom has a finite probability of being displaced with respect to its crystallographic equilibrium position. ADPs offer a convenient way to rationalize such atomic motions; for a given temperature, an ellipsoid is associated to each atom, which provides information on the probability of finding that atom displaced from the equilibrium position. A 3$\times$3 Cartesian matrix representation of the atomic ADPs can be given, according to:~\cite{NucECD,ADPMadsen}
\begin{equation}
\label{eq:adp}
B_{ij}^a(T) = \frac{1}{n_kM_a} \sum_{p{\bf k}} \frac{\mathcal{E}_{p{\bf k}}(T)}{\omega^2_{p{\bf k}}} u^{\bf k}_{ia,p} \times (u^{\bf k}_{ja,p})^\ast
\end{equation}
where the sum runs over phonons in the FBZ and $\mathcal{E}_{p{\bf k}}(T)$ is the mean vibrational energy of a phonon ({\it i.e.} harmonic oscillator) with angular frequency $\omega_{p {\bf k}}= {2\pi}\nu_{{\bf k}p}$, in thermal equilibrium at temperature $T$:~\cite{Grun_MIO}
\begin{equation}
\label{eq:ene}
\mathcal{E}_{p{\bf k}}(T) = \hbar \omega_{p{\bf k}} \left[  \frac{1}{2} + \frac{1}{e^{\frac{\hbar\omega_{p{\bf k} }}{k_B T } } - 1}  \right] \; .
\end{equation}
If ${\bf B}^a(T)$ is positive definite, then the surfaces of constant probability are defined by:
\begin{equation}
{\bf u}_a^T  {\bf B}^a(T)^{-1}  {\bf u}_a = \textup{constant}\;,
\end{equation}
consisting of \textit{ellipsoids} enclosing some finite probability for atomic displacement.~\cite{Dunitz} The lengths of the principal semi-axes of the ellipsoid and their orientations are given by the eigenvalues and eigenvectors of ${\bf B}^a(T)$, respectively. The eigenvalues $\lambda_i$ are usually expressed in units of 10$^{-4}$ \AA{}$^2$. Present calculations refer to harmonic ADPs, which correspond to a constant-volume case. In order to account for thermal changes in the cell volume, a quasi-harmonic approximation can be used instead.~\cite{Thermal_MgCaO,CORUNDUM,FORSTERITE,LIF_PRL}

\subsection{Computational Setup}
\label{comp}

All calculations were performed within unrestricted density functional theory (DFT) using the B3LYP hybrid functional,~\cite{B3LYP} as implemented in the \textsc{Crystal14} program.~\cite{CRYSTAL14PAP,cryman} A Pople's 6-21G Gaussian basis set~\cite{621G} was adopted, the exponent of the most diffuse $sp$ shell having been reoptimized in bulk diamond (0.2279 \AA$^{-2}$).
The same basis set is centered at the vacancy position to increase the variational freedom around the defect. The truncation of the Coulomb and exchange infinite series is controlled by five parameters, which were set to (8, 8, 8, 8, 16). The convergence threshold on energy for the self-consistent-field (SCF) calculations is 10$^{-8}$ Ha for structural optimization and 10$^{-9}$ Ha for vibration frequency calculations. The reciprocal space was sampled using a regular sublattice with shrinking factor of 16 (or 8, 4) for supercells containing 2 (or 8, $\geq$ 64) atoms, respectively. The number of corresponding {\bf k}-points in the irreducible part of the FBZ is 145 (or 29, 10), respectively when the full symmetry of the diamond lattice is preserved.

A supercell approach was used to simulate the neutral vacancy, where a large periodic cell is created, which is a multiple of the unit cell of the perfect system, and an atom is removed at its center to create the defect.
This scheme allows to effectively investigate the variation of any property with increasing defect concentration. Three supercells were considered containing 32, 64 and 128 atoms before vacancy creation. These cells correspond to defect densities of $5.53\times10^{21}$ cm$^{-3}$, $2.76\times10^{21}$ cm$^{-3}$ and $1.38\times10^{21}$ cm$^{-3}$, respectively and will be referred to in the following as SC$_N$ with $N$ number of atoms befor the vacancy creation.

\section{Results and Discussion}
\label{sec:resu}

\subsection{Raman spectra}

\begin{figure}[t!] 
\includegraphics[width=8.5cm]{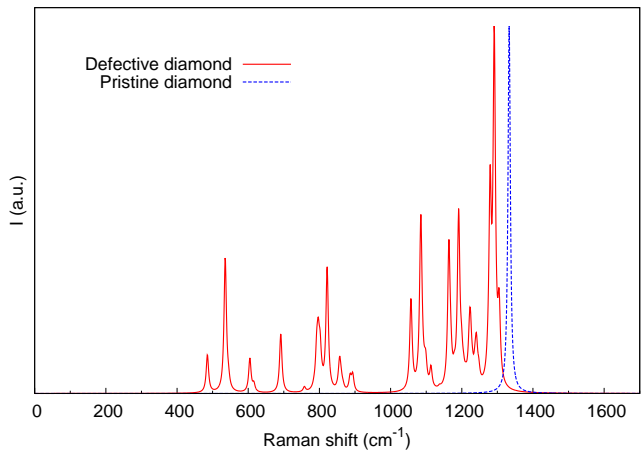} 
\caption{(color online) Raman spectra of pristine (dashed blue line) and defective (continuous red line) diamond. The $S_z=0$ spin state of the neutral vacancy in the SC$_{32}$ supercell is here considered for the defective system.}
\label{f_f1}
\end{figure}

The first-order Raman peak in pristine diamond is here found at $1333$ cm$^{-1}$, to be compared with the experimental value of $1332$ cm$^{-1}$, which confirms the effectiveness of the global hybrid B3LYP functional in describing spectroscopic features of solids. As expected, all vibration frequencies at ${\bf k} \neq \Gamma$ are found to be Raman inactive. The presence of the vacancy breaks the translational symmetry of pristine diamond and reduces the point symmetry of the system, so that new Raman-active modes emerge in the $400-1300$ cm$^{-1}$ spectral range. This effect can be clearly seen in Figure \ref{f_f1}, where we report a comparison between the Raman spectra of pristine (single peak in dashed blue line) 
and defective (spectrum in red continuous line) diamond,
modeled using a SC$_{32}$ supercell containing a vacancy in the $S_z=0$ spin state. Our results predict a large number (about 90) of Raman active phonon modes in the above-mentioned spectral range, most of which have significantly lower intensities as compared to the first-order peak.

\begin{figure}[t!]
\includegraphics[width=8.5cm]{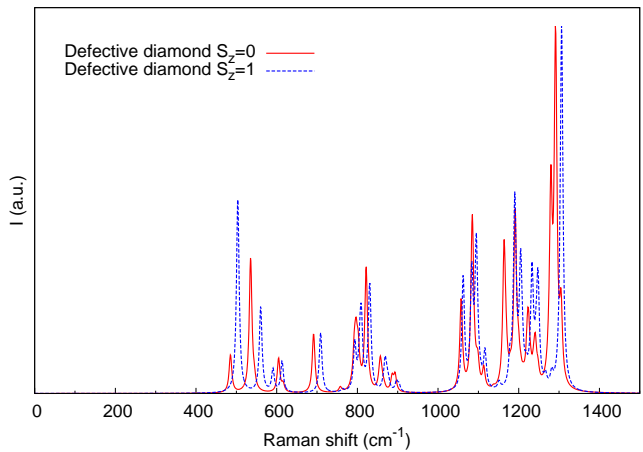} 
\caption{(color online) Raman spectra of the neutral vacancy in diamond, modeled within the SC$_{32}$ supercell, for two different spin states of the defect: $S_z=0$ (red continuous line) and $S_z=1$ (blue dashed line).}
\label{f_f2}
\end{figure}

As different spin states of the vacancy defect determine different symmetries, we investigate the effect of this feature on the Raman spectrum. 
In the present study, we consider the ground ($S_z=0$) and first excited ($S_z=1$) states of the defect. It is worth mentioning that both states involve unpaired electrons, localized on the four nearest neighbors of the vacancy, with spins in singlet and triplet configurations, respectively. It follows that open-shell calculations are performed in both cases. The two spin states are very close in energy and, thus, they both contribute significantly to the many-body wavefunction.\cite{simo-spin-vacanza}
In Figure \ref{f_f2} Raman spectra computed with the SC$_{32}$ supercell in the two spin states are reported. The main spectral features of the two cases are very similar, with slight differences in the positions and intensities of the observed peaks: i) a red-shifted first-order peak compared to pristine diamond; ii) a large number of Raman-active modes at lower frequencies, approximately in the range $400-1300$ cm$^{-1}$; iii) no peaks appearing above the first-order one. 

The analysis of the normal modes $\bf{u}^{\Gamma}_{p}$ allows to determine which spectral regions are mostly dominated by atomic motions involving nearest neighbors of the vacancy. From this analysis, we find that: i) the first-order peak only slightly involves atoms close to the vacancy; ii) the spectral region below about 750 cm$^{-1}$ is due to soft phonon modes, which strongly involve motion of those atoms, as expected; iii) an intermediate spectral region in the range 750 - 1100 cm$^{-1}$ is characterized by a small contribution from nearest neighbors; iv) in the range 1100 - 1300 cm$^{-1}$, normal modes are more affected by atoms close to the vacancy.

\begin{figure}[t!!]
\includegraphics[width=8.5cm]{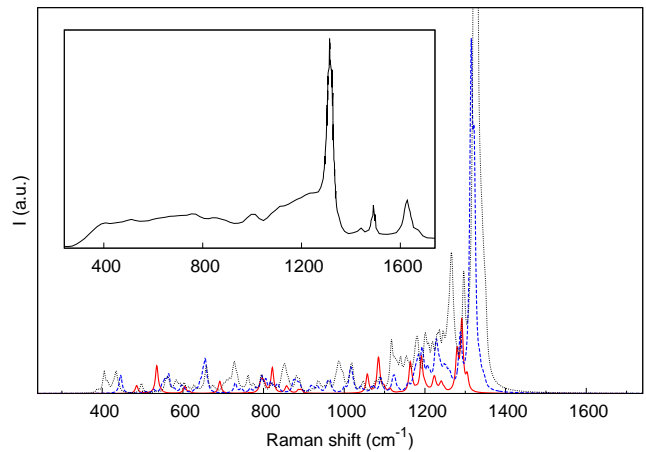} 
\caption{Raman spectra of defective diamond as a function of the density of vacancies: the continuous red line refers to the SC$_{32}$ supercell, the dashed blue line to SC$_{64}$, and the dotted black line to SC$_{128}$. The inset shows the experimental spectrum from Ref. \onlinecite{16:prawer1998}.}
\label{f_ff2}
\end{figure}

At this point, we investigate the effect of defect concentration on computed Raman spectra. Since the spin state only slightly affects the spectroscopic features of the spectrum, just the $S_z=0$ case will be considered. Three different supercells are here explored, namely SC$_{32}$, SC$_{64}$ and SC$_{128}$, which correspond to defect densities of $5.53\times 10^{21}$, $2.76\times 10^{21}$ and $1.38\times 10^{21}$ cm$^{-3}$, respectively. The three corresponding Raman spectra are reported in Figure \ref{f_ff2}, where it is shown that the effect of an increasing defect concentration is two-fold: i) a red-shift and lowering of the intensity of the first-order peak; ii) large differences in peak positions and intensities in the $400-1300$ cm$^{-1}$ region, in which local variations in defect concentration would likely determine a significant broadedning of the spectrum. Two sub-regions can be isolated in terms of the average intensity of the peaks: a first one with lower intensity between 400 and 1050 cm$^{-1}$ and a second one with higher intensity between 1050 and 1300 cm$^{-1}$. Again, no peaks are found above 1332 cm$^{-1}$.

The inset of Figure \ref{f_ff2} shows the experimental Raman spectrum of defective diamond from Ref. \onlinecite{16:prawer1998}, where the features in the $400-1300$ cm$^{-1}$ range match with the present calculations. This evidence is rather striking, particularly if it is considered that, as shown above, Raman features exhibit a spectral dispersion depending upon the particular spin state of the defect, with frequency shifts down to few wave-numbers, which might be hard to discriminate with the spectral resolution of common experimental setups. %Moreover, it is worth remarking that in the present work we did not compute the excited $S_z=2$ spin state, which most certainly would introduce additional slight shifts of Raman features in the same spectral ranges, thus contributing to the overall broadening of the spectrum. 

Apart from concentration, two further issues may affect the comparability between present calculations and experimental data: 
i) the defective system is here modeled as a periodic repetition of point defects with a spectral dispersion caused by the variation in the defect concentration, while in experimental conditions a random spatial distribution of point defects is obtained with all techniques employed to induce structural damage; 
ii) our model is here exclusively limited to isolated vacancy defects, while a broad variety of point (or more complex) defects is created in diamond upon ion implantation due to its peculiar meta-stability.\cite{22:zaitsev2001} 
Taking into account the above-mentioned limitations of the model, the agreement between present quantum-mechanical calculations and experimental measurements constitutes a strong indication that isolated vacancies do play a prominent role in determining the spectral features of defective diamond below $1332$ cm$^{-1}$, thus confirming the hypothesis that Raman features in the $400-1300$ cm$^{-1}$ range should be mainly attributed to non-graphitic sp$^3$ defects. \cite{16:prawer1998}

On the other hand, by comparing computed with experimental spectra in Figure \ref{f_ff2}, it is apparent that no Raman features are found in the simulated spectrum above $1332$ cm$^{-1}$, differently from the experimental one where several features are measured at about 1450, 1490, 1630 and 1680 cm$^{-1}$. This evidence clearly indicates that the two experimentally measured sharp features at about 1450 and 1490 cm$^{-1}$ in defective diamond cannot be attributed, as previously suggested, \cite{16:prawer1998,20:prawer2008} to isolated vacancies but rather to other kinds of defects involving, for instance, intrinsic or nitrogen interstitials.\cite{24:woods1984, 25:linchung1994}

\begin{figure}[t!!]
\includegraphics[width=8.5cm]{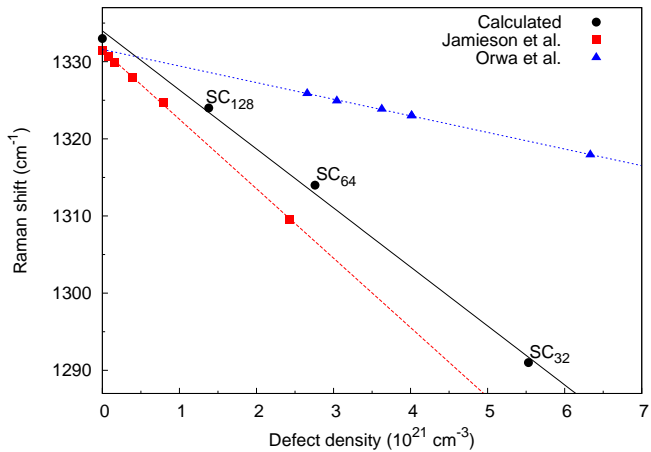} 
\caption{(color online) First-order Raman peak position of defective diamond as a function of defect density. Computed values (black circles) are compared with experimental data by Jamieson {\it et al.}\cite{14:jamieson1995} (red squares) and by Orwa {\it et al.}\cite{17:orwa2000} (blue triangles). Lines are linear fits of the data (see text for details).}
 \label{f_ff3}
\end{figure}

In order to further confirm the reliability and predictiveness of the present model, a quantitative comparison between computed and experimental results is presented on the position of the first-order peak in the Raman spectrum as a function of defect concentration. A quasi-linear red-shift of that peak is indeed observed as the defect density increases. Present computed data are reported in Figure \ref{f_ff3} (black circles) and compared with experimental data by Jamieson {\it et al.}\cite{14:jamieson1995} (red squares) and by Orwa {\it et al.}\cite{17:orwa2000} (blue triangles). In the former experimental work,\cite{14:jamieson1995} the Raman red-shift was measured in a cross-sectional geometry at $1\ \mu$m depth in single-crystal diamond implanted with 3.5 MeV He$^{+}$ ions at increasing fluences ranging from $1\times10^{16}$ cm$^{-2}$ to $3\times10^{17}$ cm$^{-2}$. In the present work, those fluence values (\textit{i.e.} number of implanted ions per surface unit) are converted into volumetric density of vacancies by multiplying by the average linear density of vacancies created per incoming ion at the above-mentioned depth, as derived from Monte Carlo simulations by use of the SRIM Monte Carlo code.\cite{11:zie} %SRIM ver 2013.00  
Similarly, in the latter experimental work,\cite{17:orwa2000} cross-sectional micro-Raman measurements were performed at different depths (\textit{i.e.} probing different damage densities) in a single-crystal diamond implanted with 3 MeV He$^{+}$ ions at a fixed fluence of $1\times10^{17}$ cm$^{-2}$. In this case, vacancy densities values were already derived from SRIM simulations. In Figure \ref{f_ff3}, linear fits of computed and experimental data are also reported. The slope of our computed data, (-7.6$\pm$0.3)$\times 10^{-21}$ cm$^{3}$, is comprised between that of Jamieson {\it et al.},\cite{14:jamieson1995} (-9.0$\pm$0.1)$\times 10^{-21}$ cm$^{3}$, and that of Orwa {\it et al.},\cite{17:orwa2000} (-2.2$\pm$0.3)$\times 10^{-21}$ cm$^{3}$.

We observe that present theoretical results exhibit a better consistency with the experimental results obtained by the cross-sectional Raman probing the damaged structure at a fixed depth of 1 $\mu$m under increasing implantation fluences,\cite{14:jamieson1995} with respect to similar measurements performed on a sample implanted at a single fluence at different depths.\cite{17:orwa2000} This is surprising, given the fact that the same ions (\textit{i.e.} He$^{+}$) were implanted at similar energies (\textit{i.e.} 3.5 MeV and 3 MeV, respectively) in the two above-mentioned experiments. We interpret the better adherence of the theoretical results with the results obtained in Ref. \onlinecite{14:jamieson1995} by considering that the employment of multiple implantations should minimize the effects of the uncertainty of an important implantation parameter, \textit{i.e.} the fluence, thus reducing the probability of possible systematic errors.
Overall, the good consistency between the reported computed data and the experimental results, despite the significant approximation of modeling the defective system as a regular pattern, represents a further indication that isolated vacancies play a major role in determining the measured Raman features of defective diamond.

\subsection{Anisotropic Displacement Parameters}

As anticipated in Section \ref{subsec:adp}, the computation of atomic ADPs represents an effective way to interpret thermal nuclear motions, allowing the quantification of the mean square atomic displacements in an intuitive way. Indeed, with respect to vibration normal coordinates, which are collective modes, ADPs allow for an atomic partition of the thermal nuclear motion, providing a simple graphical tool to study these effects, even in complex structures. Furthermore, ADPs can be experimentally derived from X-ray or synchrotron radiation diffraction measurements, thus allowing for validation of the theoretical predictions. The atomic thermal ellipsoids of pure diamond are perfectly spherical and the corresponding ADP has a value which is commonly reported in the range (16.1 - 22.3)$\times$10$^{-4}$\AA$^2$, with a recent accurate determination of 18.1$\times$10$^{-4}$\AA$^2$.\cite{ADP_DIAM} The {\it ab initio} evaluation of ADPs requires lattice dynamical calculations to be performed by accounting for phonon dispersion. Within the direct space (frozen phonon) approach adopted here, increasing the size of the SC in the lattice dynamical calculation corresponds to increasing the sampling of the phonon dispersion within the first Brillouin zone in reciprocal space. 

\begin{table}[t!!]
%\large
\centering
\caption{Computed ADP (in units of 10$^{-4}$\AA$^2$) of pure crystalline diamond at 298.15 K as a function of the size of the SC used for the lattice dynamical calculations, $N_\textup{at}$ being the number of atoms in the SC. The corresponding number $n_k$ of {\bf k}-points over which the FBZ is sampled is also given. A recent experimental determination is reported as a reference.\cite{ADP_DIAM}} 
\label{tab:adp}
\begin{tabular}{lcr}
\hline
\hline
\vspace{2pt}$N_\textup{at}$ & $n_k$& ADP \\ \hline
2 & 1& 4.1 \\
8 & 4 & 11.7 \\
16 & 8 & 14.2 \\
32 & 16& 14.5 \\
64 & 32& 15.3 \\
128 & 64& 15.9 \\
216 & 108& 16.4 \\
\vspace{4pt}256 & 128& 16.5 \\
\vspace{4pt} Interp. & 65536 & 17.7 \\
\vspace{2pt} Exp. && 18.1\\
\hline \hline
\end{tabular}
\end{table}

The computed ADP of pure diamond at 298.15 K, as obtained by use of equation (\ref{eq:adp}), is reported in Table \ref{tab:adp} as a function of the size of the adopted SC. The effect of the inclusion of phonon dispersion is rather large: a $\Gamma$-only calculation (performed on the primitive unit cell containing only $N_\textup{at}$ = 2 atoms) indeed provides an ADP of 4.1 10$^{-4}$\AA$^2$, which is then systematically increased as the size of the SC increases. Convergence is reached for a SC containing $N_\textup{at}$ = 256 atoms (corresponding to a sampling of the FBZ over 128 {\bf k}-points), as it provides a value of 16.5$\times$10$^{-4}$\AA$^2$ compared to 16.4$\times$10$^{-4}$\AA$^2$ that is obtained with a SC with $N_\textup{at}$ = 216 atoms. Without further increasing the size of the SC, a denser sampling of phonon dispersion can then be achieved by Fourier interpolation of the dynamical matrices. Indeed, if equation (\ref{eq:dynmat}) could be in principle used to compute and then diagonalize the dynamical matrices of just the $L = \prod_i L_i$ {\bf k}-points defined in Section \ref{subsec:phon}, this restriction would disappear when long-range electrostatic contributions to the force constants vanish within the SC (as in the case of diamond). In this case, such an expression can be used to construct the dynamical matrices of a denser set of {\bf k}-points through Fourier interpolation. The ADP obtained by Fourier interpolation (with a corresponding sampling of phonon dispersion over 65,536 {\bf k}-points), starting from the largest SC, is 17.7$\times$10$^{-4}$\AA$^2$, which is in satisfactory agreement with the accurate experimental reference of 18.1$\times$10$^{-4}$\AA$^2$,\cite{ADP_DIAM} thus showing the accuracy of the convergence of the lattice dynamical description of the system. 

\begin{figure}[t!!]
\centering
\includegraphics[width=8.5cm]{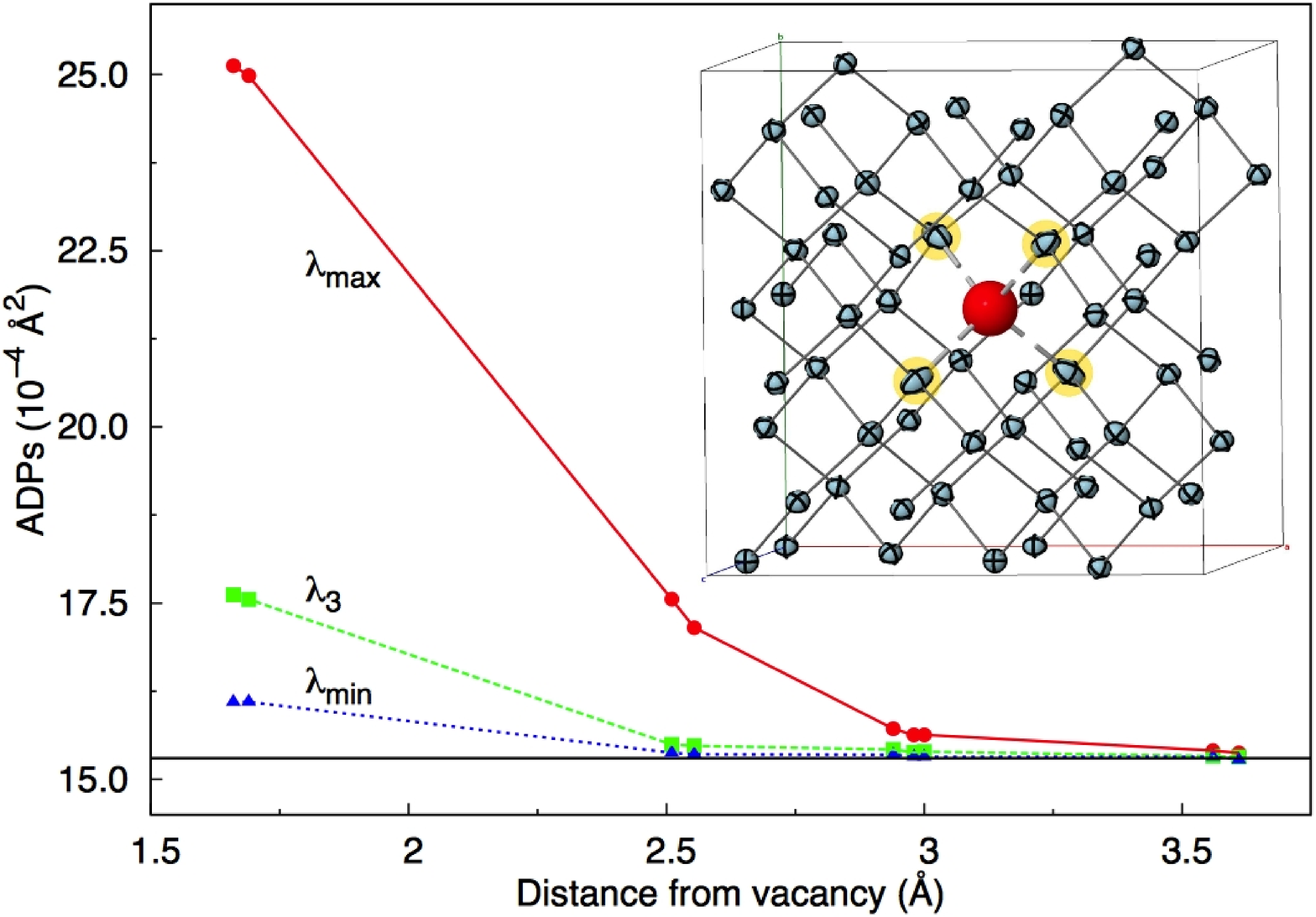}
\caption{(color online) The three semi-axes ($\lambda_{\mathrm{max}}$, $\lambda_{\mathrm{min}}$ and $\lambda_3$) of the ADP ellipsoids of carbon atoms as a function of their distance from the vacancy. The calculation is performed at 298.15 K in the SC containing 64 centers (\textit{i.e.} 63 carbons and 1 vacancy). The horizontal black line marks the value of the isotropic ADP of pure diamond, as obtained from lattice dynamical calculations performed on a SC of the same size. In the inset, a graphical representation is reported of the ADPs ({\it i.e.} thermal ellipsoids) relevant to the 63 carbon atoms surrounding the vacancy (in red) in the SC$_{64}$ cell. Nearest neighbors of the vacancy are highlighted in yellow.}
\label{fig:adp2}
\end{figure}

We shall now analyze the atomic ADPs in defective diamond, \textit{i.e.} the size, shape and orientation of the thermal ellipsoids of carbon atoms surrounding the vacancy in the lattice. As the vacancy constitutes a local perturbation to the lattice, the features of the ADPs in the vicinity of the defect are found to be almost independent from defect concentration (at least at low defect concentrations). For this reason, we restrict our analysis to the SC containing 64 centers (\textit{i.e.} 63 carbons and 1 vacancy). In the inset of Figure \ref{fig:adp2}, a graphical representation is given of all thermal ellipsoids associated with the 63 carbon atoms of the SC. ADPs of atoms sufficiently far apart from the vacancy (represented as a red sphere at the body center of the cell) are almost perfectly isotropic, as in the ideal non-perturbed diamond lattice. When carbon atoms close to the vacancy are considered (particularly for the four nearest neighbors, highlighted in yellow in the figure), the picture is rather different, as the ellipsoids are quite elongated along the axis connecting each atom to the center of the vacancy. In other words, atoms close to the vacancy do exhibit a larger thermal mobility towards the vacancy, as expected.

In order to determine the maximum distance over which the thermal motion of the atoms is affected by the vacancy, we refer to Figure \ref{fig:adp2}. The lengths of the three semi-axes of the thermal ellipsoids ($\lambda_\textup{max}$, red circles, $\lambda_\textup{min}$, blue triangles, and $\lambda_3$, green squares) are reported as a function of the distance of each carbon atom from the vacancy. Atoms close to the vacancy show a highly anisotropic thermal ellipsoid, with the longest ADP, $\lambda_\textup{max}$, almost 56\% larger than the shortest one, $\lambda_\textup{min}$. Second nearest neighbors (at a distance of about 2.5 \AA) are still affected by the vacancy while atoms at distances larger than 3 \AA\ are practically not affected by the vacancy, thus showing a recovered isotropic character of the ADP. 
%The black horizontal line marks the isotropic reference value of the ADP of pure diamond with the same SC.   

\section{Conclusions}
\label{sec:concl}

Quantum-mechanical {\it ab initio} calculations were performed to investigate Raman spectroscopic features of defective diamond due to the neutral vacancy point-defect. Raman spectra were computed analytically, through a Coupled-Perturbed-Hartree-Fock/Kohn-Sham approach as a function of both defect spin state and concentration, by means of a supercell periodic model. Experimentally observed features in the Raman spectrum of irradiation-damaged diamond are well reproduced for Raman shifts below 1332 cm$^{-1}$ (\textit{i.e.} to the first-order peak of pristine diamond), thus supporting the attribution of these features to non-graphitic sp$^3$ defects. No spectral features are predicted at frequencies above 1332 cm$^{-1}$, an evidence that rules out previous tentative assignments of the experimentally measured peaks at about 1450 and 1490 cm$^{-1}$ to the neutral vacancy.

The thermal nuclear motion of carbon atoms in both the perfect and defective diamond lattice was also investigated. Atomic anisotropic displacement parameters (ADPs) were computed from converged phonon dispersion calculations. The thermal motion of the first neighbors of the vacancy was found to be significantly affected by the defect, yielding a thermal ellipsoid which is rather elongated towards the defect (the principal axis being almost twice as long as the other ones). At room temperature, the spherical shape of the ADP (typical of pristine diamond) is fully recovered at a distance of about 3.5 \AA{} from the vacancy.

% \bibliography{biblio,DatabaseCRYSCOR2,DatabaseCRYSCOR,other_articles}

\bibliography{}

\end{document}